\documentclass[aps,prb,amsmath,amsfonts,superscriptaddress,twocolumn,preprintnumbers,floatfix,longbibliography]{revtex4-1}
\usepackage{amssymb}
\usepackage{graphicx}
\usepackage{appendix}
\usepackage{color}
\usepackage{subfigure}
\usepackage{float}
\usepackage[colorlinks=true,linkcolor=blue,filecolor=blue, urlcolor=blue,citecolor=blue]{hyperref}
\allowdisplaybreaks[4]

\begin{document}
% Use the \preprint command to place your local institutional report
% number in the upper righthand corner of the title page in preprint mode.
% Multiple \preprint commands are allowed.
% Use the 'preprintnumbers' class option to override journal defaults
% to display numbers if necessary
%\preprint{}

\title{An improved particle swarm optimization algorithm and its application to search for new magnetic ground states in the Hubbard model}

% repeat the \author .. \affiliation  etc. as needed
% \email, \thanks, \homepage, \altaffiliation all apply to the current
% author. Explanatory text should go in the []'s, actual e-mail
% address or url should go in the {}'s for \email and \homepage.
% Please use the appropriate macro foreach each type of information

% \affiliation command applies to all authors since the last
% \affiliation command. The \affiliation command should follow the
% other information
% \affiliation can be followed by \email, \homepage, \thanks as well.
\author{Ze Ruan}
%\homepage[]{Your web page}
%\thanks{}
%\altaffiliation{}
\email{These authors contribute equally to the paper.}
\affiliation{Shanghai Key Laboratory of Special Artificial Microstructure Materials and Technology, School of Physics Science and engineering, Tongji University, Shanghai 200092, P.R. China}

\author{ Xiu-Cai Jiang}
\email{These authors contribute equally to the paper.}
\affiliation{Shanghai Key Laboratory of Special Artificial Microstructure Materials and Technology, School of Physics Science and engineering, Tongji University, Shanghai 200092, P.R. China}

\author{Ze-Yi Song}
\affiliation{Shanghai Key Laboratory of Special Artificial Microstructure Materials and Technology, School of Physics Science and engineering, Tongji University, Shanghai 200092, P.R. China}

\author{Yu-Zhong Zhang}
\email[Corresponding author: ]{yzzhang@tongji.edu.cn}
\affiliation{Shanghai Key Laboratory of Special Artificial Microstructure Materials and Technology, School of Physics Science and engineering, Tongji University, Shanghai 200092, P.R. China}
%Collaboration name if desired (requires use of superscriptaddress
%option in \documentclass). \noaffiliation is required (may also be
%used with the \author command).
%\collaboration can be followed by \email, \homepage, \thanks as well.
%\collaboration{}
%\noaffiliation

\date{\today}

\begin{abstract}
An improved particle swarm optimization algorithm is proposed and its superiority over standard particle swarm optimization algorithm is tested on two typical benchmark functions. By employing this algorithm to search for the magnetic ground states of the Hubbard model on the real-space square lattice with finite size based on the mean-field approximation, two new magnetic states, namely the double striped-type antiferromagnetic state and the triple antiferromagnetic state, are found. We further perform mean-field calculations in the thermodynamical limit to confirm that these two new magnetic states are not a result of a finite-size effect, where the properties of the double striped-type antiferromagnetic state are also presented.
\end{abstract}

\maketitle

\section{Introduction}
Antiferromagnetism has attracted tremendous interests due to the fact that the parent states of cuprates and iron-based superconductors are either antiferromagnets or paramagnets with antiferromagnetic fluctuations, such as the checkerboard antiferromagnetic (CAF) insulator of cuprates, nearly degenerate double-stripe (DAF) and plaquette antiferromagnetic (PAF) order in FeTe\cite{li2009first,tam2019plaquette}, pair-checkerboard antiferromagnetic (PCAF) order in monolayer FeSe thin film\cite{zhou2018antiferromagnetic,cao2015antiferromagnetic}, molecular-intercalated FeSe\cite{taylor2013spin}, and A$_{x}$Fe$_{2-y}$Se$_{2}$\cite{taylor2012spin}, as well as stripe-type antiferromagnetic (SAF) order in other iron-based mother materials\cite{de2008magnetic,huang2008neutron,li2009structural}, which indicates a magnetic origin of the two high-$T_{c}$ unconventional superconductors.
\par
Therefore, much effort has been spent on studying the properties of these magnetic states based on models related to these two superconducting families. A simple Heisenberg model with the nearest and next-nearest neighbour intralayer couplings has been used to explain the transition from CAF to SAF~\cite{chandra1990ising}. Additionally, a combination of this simple Heisenberg model with the third-neighbour intralayer coupling solely~\cite{ma2009first,ducatman2012magnetism}, or with three couplings~\cite{hu2012unified}, namely the third-neighbour intralayer coupling, nearest interlayer coupling, and intralayer nearest-neighbour biquadratic coupling, is introduced to further include DAF and PAF. In addition to these models, an effective orbital-degenerate double-exchange model consisting of both itinerant electrons and localized spins is used to take CAF, DAF, and SAF into consideration~\cite{yin2010unified}. Alternatively, the Hubbard model which covers both weak and strong coupling limits for magnetism is also employed to investigate these magnetic states, for example, the single-orbital Hubbard model with the nearest and next-nearest neighbour hoppings. Using methods like variational cluster approximation~\cite{yamaki2013ground,nevidomskyy2008magnetism}, variational Monte Carlo\cite{tocchio2008role}, and mean-field theory~\cite{yu2010collinear}, the presence of CAF and SAF in this model is proposed. Apart from these two states, PCAF is also found in this model by path-integral renormalization group method~\cite{mizusaki2006gapless} and variational cluster approximation~\cite{yamada2013magnetic}. Recently, it has been pointed out that this Hubbard model can serve as a unified minimal model to describe all the magnetic states mentioned above~\cite{ruan2021uncovering}.
\par
However, despite numerous investigations, only a few new magnetic states which probably exist in the experiment are predicted based on the aforementioned superconductor-related magnetic models. For example, the spiral and staggered trimer states are predicted in $J_1$-$J_2$-$J_3$-$K$ Heisenberg model~\cite{glasbrenner2015effect}, or a spiral state is found in a frustrated Hubbard model containing the nearest and next-nearest neighbour hoppings, where the latter hopping breaks $C_4$ symmetry~\cite{misumi2016phase}. The lack of new magnetic states may be due to the weakness of conventional methods, such as variational cluster approximation, variational Monte Carlo, mean-field theory, etc., as they require preparation of the desired configurations while it is impossible to exhaust all magnetic patterns. Thus, a natural question arises: whether more new exotic magnetic states occur in the aforementioned unified minimal Hubbard model using a method superior to conventional methods? Noticeably, the particle swarm optimization (PSO) algorithm is proven to be a powerful approach~\cite{kennedy1995particle,eberhart1995new}, which is also used in the prediction of new materials~\cite{wang2010crystal,zhang2014robust,qu2022particle,yan2023designing} and the estimation of cosmological parameters~\cite{prasad2012cosmological,ruiz2015calibration}, etc.
Regarding that PSO can handle problems with numerous energy minima~\cite{wang2010crystal} and generate random particles, it may be suitable to search for new magnetic states, where a magnetic pattern is viewed as a particle. Nevertheless, the insufficient performance of standard PSO that sometimes converges slowly~\cite{beheshti2014capso} or tends to converge to local minima~\cite{clerc2002particle} impedes such application. Although lots of attempts have been undertaken to improve the performance of the standard PSO~\cite{mendes2004fully,chatterjee2006nonlinear,wang2011self,engelbrecht2012particle,li2015competitive,munlin2017new,wu2019denpso,jakubik2021directed}, an improved PSO with the hyperparameters (including the inertia weight and learning factors) depending on individual character of particle has not yet been proposed. Therefore, proposing an improved PSO with particle-dependent hyperparameters and using it to search for new magnetic states in the aforementioned Hubbard model is an interesting work.
\par
In this paper, we proposed an improved PSO, where the hyperparameters of a specific particle depend on its present position, corresponding local best, and global best. We first show the superiority of this improved PSO over standard PSO on two typical benchmark functions, namely the Griewank function and Rastrigrin function. Then, by employing this improved PSO to search for the magnetic ground state of the Hubbard model on the real-space square lattice with a finite size of 24$\times$24 based on the mean-field approximation, two new magnetic states are found, namely the double stripe-type state (DSAF) and the triple antiferromagnetic state (TAF). By comparing the free energies of these two new states with their competing states in corresponding parameter space using mean-field theory in the thermodynamical limit, we further confirm that these two new magnetic states are not a result of a finite-size effect. Besides, we present the properties of DSAF which occurs at a region of weak frustration.
\par
Our paper is organized as follows. Section \uppercase\expandafter{\romannumeral2} describes the Hubbard model and the standard PSO. Section \uppercase\expandafter{\romannumeral3} presents our main results, including the superiority of our improved PSO over the standard PSO, the search of new magnetic ground states of Hubbard model on the square lattice with a finite real-space size using improved PSO, and the mean-field calculations in the thermodynamical limit. Section \uppercase\expandafter{\romannumeral4} presents a detailed discussion, and Section \uppercase\expandafter{\romannumeral5} concludes with a summary.

\section{model and method}
The Hubbard model on a square lattice with the nearest and next-nearest neighbour hoppings we studied is given by
\begin{small}
\begin{equation}
H=-t_{1}\sum_{\langle i,j\rangle,\sigma}c^{\dagger}_{i\sigma}c_{j\sigma}
-t_{2}\sum_{\langle\langle i,j\rangle\rangle,\sigma}c^{\dagger}_{i\sigma}c_{j\sigma}
+U\sum_{i}n_{i\uparrow}n_{i\downarrow},
\end{equation}
\end{small}
where $c^{\dagger}_{i\sigma}$($c_{i\sigma}$) creates (annihilates) an electron at site i with spin $\sigma$. $n_{i\sigma}$ is the number operator. $t_{1}$ and $t_{2}$ denote the nearest- and next-nearest-neighbour hoppings, and $U$ is the on-site Coulomb interaction. $\langle i,j\rangle$($\langle\langle i,j\rangle\rangle$) means the summation over nearest (next-nearest)-neighbor sites. The on-site Coulomb interaction is treated by mean-field approximation as
\begin{equation}
Un_{i\uparrow}n_{i\downarrow}\approx Un_{i\uparrow}\langle n_{i\downarrow}\rangle +U\langle n_{i\uparrow}\rangle n_{i\downarrow}-U\langle n_{i\uparrow}\rangle\langle n_{i\downarrow}\rangle.
\end{equation}
By defining respectively the magnetic moment and charge occupation of $i$ site as $m_{i}=\langle n_{i\uparrow}\rangle-\langle n_{i\downarrow}\rangle$ and $\bar{n}_i=\langle n_{i\uparrow}\rangle+\langle n_{i\downarrow}\rangle$, we have
\begin{equation}
\langle n_{i\uparrow}\rangle=\frac{1}{2}(\bar{n}_i+m_{i}),\quad\langle n_{i\downarrow}\rangle=\frac{1}{2}(\bar{n}_i-m_{i}).
\end{equation}
Thus, for a given charge distribution and magnetic configuration $\{\bar{n}_1,\cdots,\bar{n}_N,m_1,\cdots,m_{N}\}$, the free energy can be calculated as
\begin{align}
F=-\frac{1}{\beta}\rm{ln}\Xi+\sum\limits_{i=1}^{N}\big(\mu+\frac{\emph{U}\emph{m}_\emph{i}^2}{4}-\frac{\emph{U}\bar{\emph{n}}_\emph{i}^2}{4}\big)
\label{free-energy},
\end{align}
in which the grand partition function reads
\begin{align}
\rm{ln}\Xi=\sum\limits_{\sigma}\sum\limits_{i=1}^{N}\rm{ln}\Big[1+e^{-\beta(\emph{E}_{\emph{i}\sigma}-\mu)}\Big],
\label{partition-func}
\end{align}
where $N$ is the total number of sites in the system, $\beta$ is the inversed temperature defined as $1/(k_BT)$, $\mu$ denotes the chemical potential, and $E_{i\sigma}$ is the eigenvalue derived by diagonalizing the Hamiltonian matrix of the system. Specifically, for the case with uniform charge distribution at half-filling, namely $\bar{n}_1=\cdots=\bar{n}_N=1$, the free energy is a function of the magnetic configuration $\{m_1,\cdots,m_{N}\}$.
\par
\begin{figure}
\includegraphics[width=0.38\textwidth,height=0.30\textwidth]{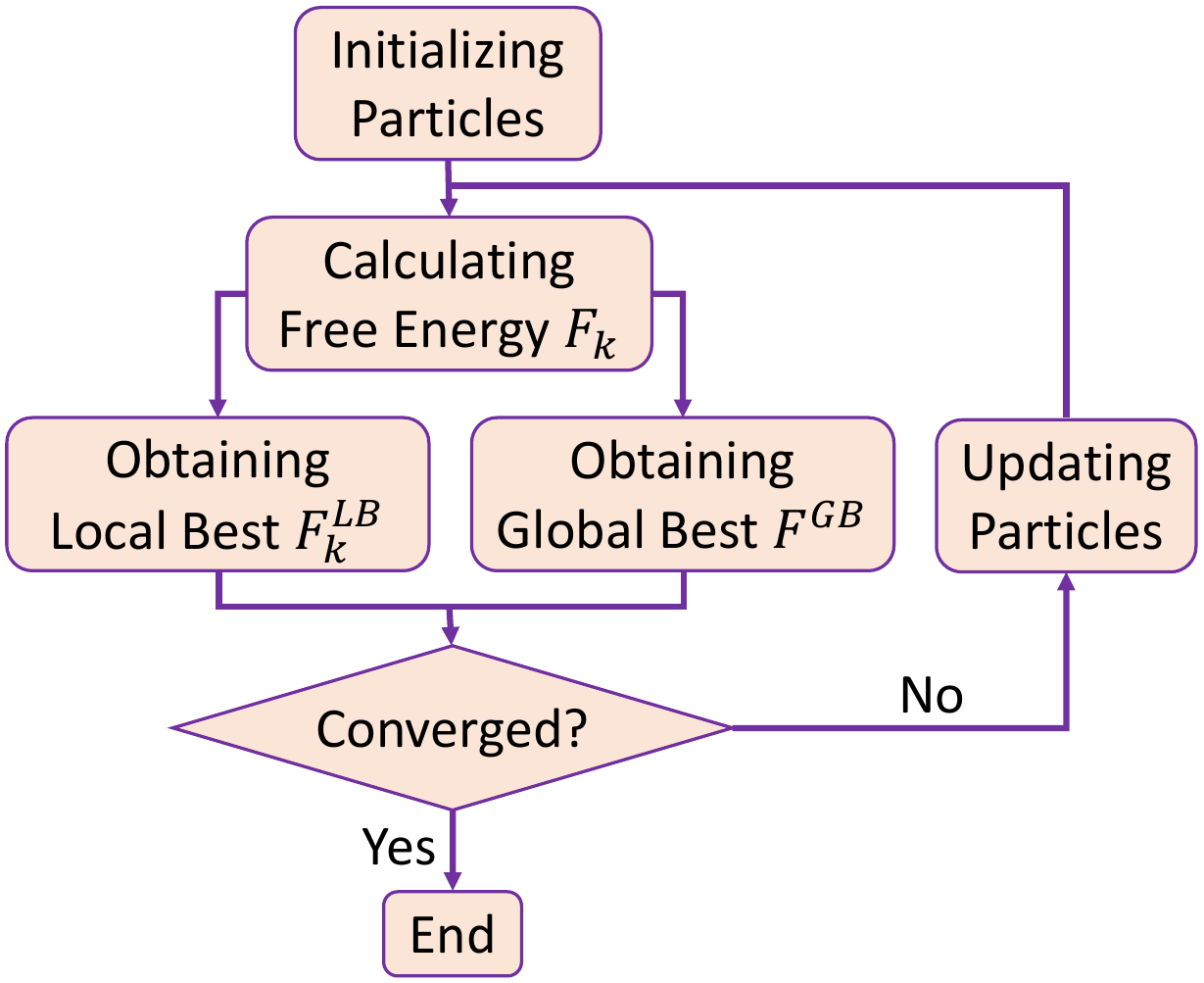}
\caption{The flowchart of PSO, where local best and global best are the minimal free energy of $k$-th particle and all particles, respectively. When the convergence criterion is reached, $|F_k^{LB}-F^{GB}|<\epsilon$ satisfies for any given $k$-th particle, where $\epsilon$ is a small positive number.}
\label{flowchart}
\end{figure}
We proceed to demonstrate how this Hubbard model is solved within the framework of PSO based on the mean-field approximation, where a magnetic configuration is viewed as a particle and free energy $F$ is the target function to be optimized to a minimum value. The flowchart of PSO is shown in Fig.\ref{flowchart}. As can be seen, the first iteration consists of following 4 steps while other iterations contain the last 3 of these steps within the framework of PSO:

(\romannumeral1) Initializing randomly particles, namely a set of magnetic configurations $\{X_1,\cdots,X_{k},\cdots,X_{P}\}$, where $X_{k}=\{m_1^k,\cdots,m_{N}^k\}$ and $P$ is the total number of particles.

(\romannumeral2) Calculating the free energies of these magnetic configurations $\{F_1,\cdots,F_{k},\cdots,F_{P}\}$ according to Eq.\eqref{free-energy}.

(\romannumeral3) Evaluating local best $\{F_1^{LB},\cdots,F_{k}^{LB},\cdots,F_{P}^{LB}\}$ and global best $F^{GB}$ based on $\{F_1,\cdots,F_{k},\cdots,F_{P}\}$ of current and last iterations.

(\romannumeral4) Determining whether the convergence criterion is reached. If $|F_k^{LB}-F^{GB}|<\epsilon$ satisfies for any given $k$-th particle, where $\epsilon$ is a small positive number, the calculation is completed, otherwise, updating the magnetic configurations and turning to step (\romannumeral2) using the following equations
\begin{footnotesize}
\begin{eqnarray}
\left\{
\begin{aligned}
D_{k}^{New}&=\omega D_{k}^{Old}+c_1r_1(X_{k}^{LB}-X_{k})+c_2r_2(X^{GB}-X_{k})\\
X_{k}^{New}&=X_{k}+D_{k}^{New}
\end{aligned}
\right.,
\label{PSO-updating}
\end{eqnarray}
\end{footnotesize}
where $\omega$ denotes the inertia weight, $c_1$ and $c_2$ are the learning factors, and $r_1$($r_2$) is a random number between 0 and 1 that vary at each iteration. $X_{k}^{LB}$ and $X^{GB}$ are the magnetic configurations of local best and global best, respectively. $D_{k}^{New}$ and $D_{k}^{Old}$ denote separately the displacement between the new and current k-th magnetic configurations, and that between the current and last k-th magnetic configurations, in which $D_{k}^{Old}$ should be initialized at the first iteration. It is  necessary to mention that, for standard PSO, $c_1$, $c_2$, and $\omega$ are  particle-independent, where $c_1$ and $c_2$ are two constants while $\omega$ decreases linearly at first and then remains a small value with the proceeding of iteration.

In contrast to standard PSO, the learning factors and inertia weight of our improved PSO are particle-dependent, where $c_1$ and $c_2$ have the forms
\begin{eqnarray}
c_{1k}=\frac{e^{F^{GB}}}{e^{F^{GB}}+e^{F_k^{LB}}},\quad
c_{2k}=\frac{e^{F_k^{LB}}}{e^{F^{GB}}+e^{F_k^{LB}}},
\label{c1c2-improved}
\end{eqnarray}
while $\omega$ reads
\begin{footnotesize}
\begin{eqnarray}
\omega_{k}=
\left\{
\begin{aligned}
&\xi\quad\quad\quad\quad\quad\quad\quad\quad\quad\quad\quad(\emph{F}_{\emph{k}}\leqslant \emph{F}^{\emph{GB}}\cap \emph{F}_{\emph{k}}\leqslant \emph{F}_\emph{k}^{\emph{LB}})\\
&(F_{k}-F^{GB})^{\gamma}\quad\quad\quad\quad\quad(F^{GB}<F_{k}<F_k^{LB})\\
&(2F_{k}-F^{GB}-F_k^{LB})^{\gamma}\quad(F^{GB}<F_{k}\cap{F}_k^{LB}<F_{k})
\end{aligned}
\right.,
\label{omega-improved}
\end{eqnarray}
\end{footnotesize}
in which $\gamma$ is a tunable positive number with the highest performance when $1.5\sim1.7$ for Rastrigrin function, $5\sim7$ for the Griewank function, and 0.01 for this Hubbard model while $\xi$ is a small positive number. Thus, by combining equations \eqref{free-energy}-\eqref{omega-improved}, the Hubbard model can be solved self-consistently using our improved PSO.

\section{results}
\subsection{The superiority of our improved PSO over standard PSO}
\begin{figure}
\centering
\includegraphics[width=0.36\textwidth,height=0.54\textwidth]{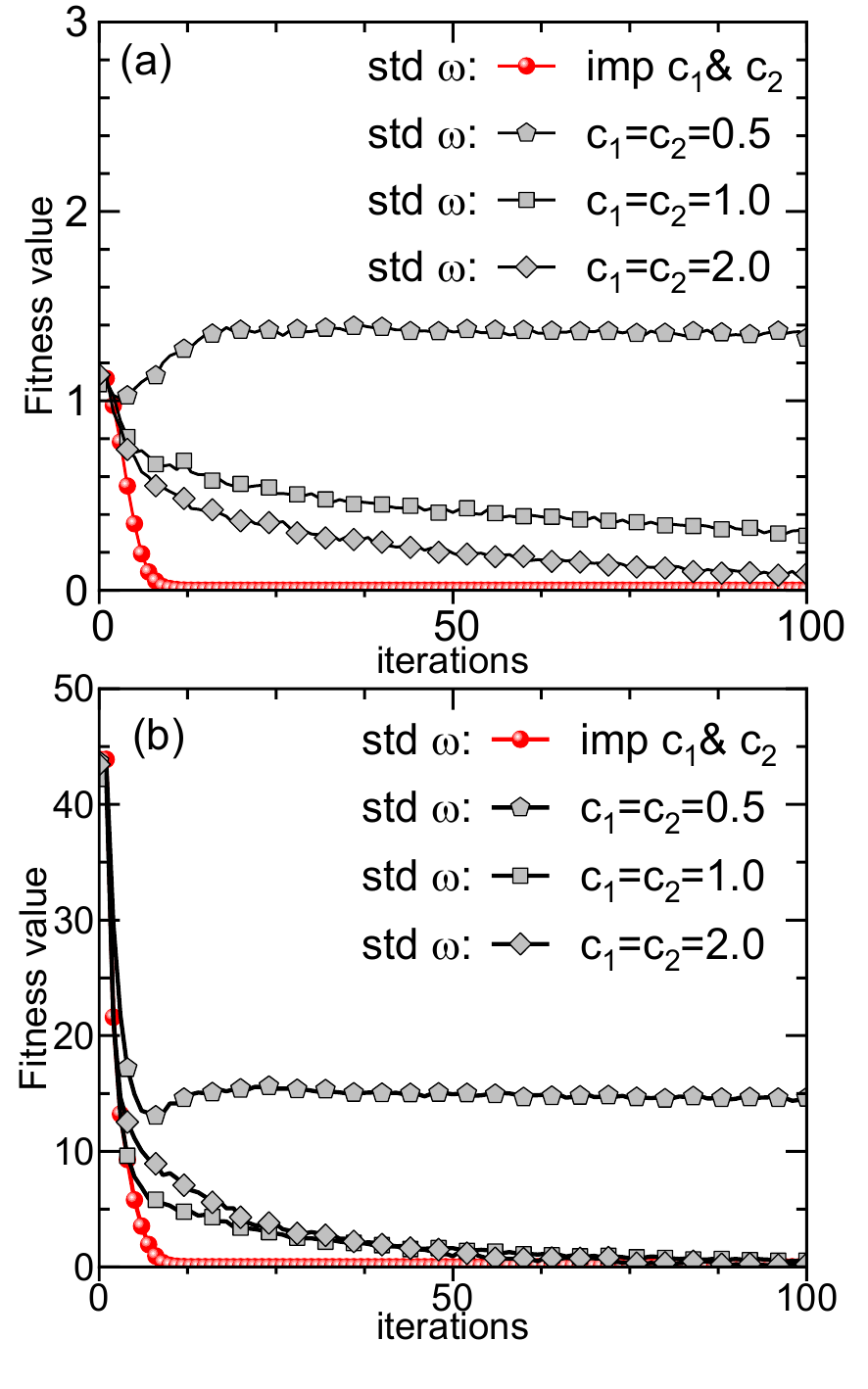}
\caption{The convergence process on the Griewank function (a) and Rastrigrin function (b) using PSO with improved $c_1$ and $c_{2}$ or with standard $c_{1}$ and $c_{2}$ where the inertia weight ($\omega$) of standard PSO is used.}
\label{learningfactors}
\end{figure}
\begin{figure}
\centering
\includegraphics[width=0.36\textwidth,height=0.54\textwidth]{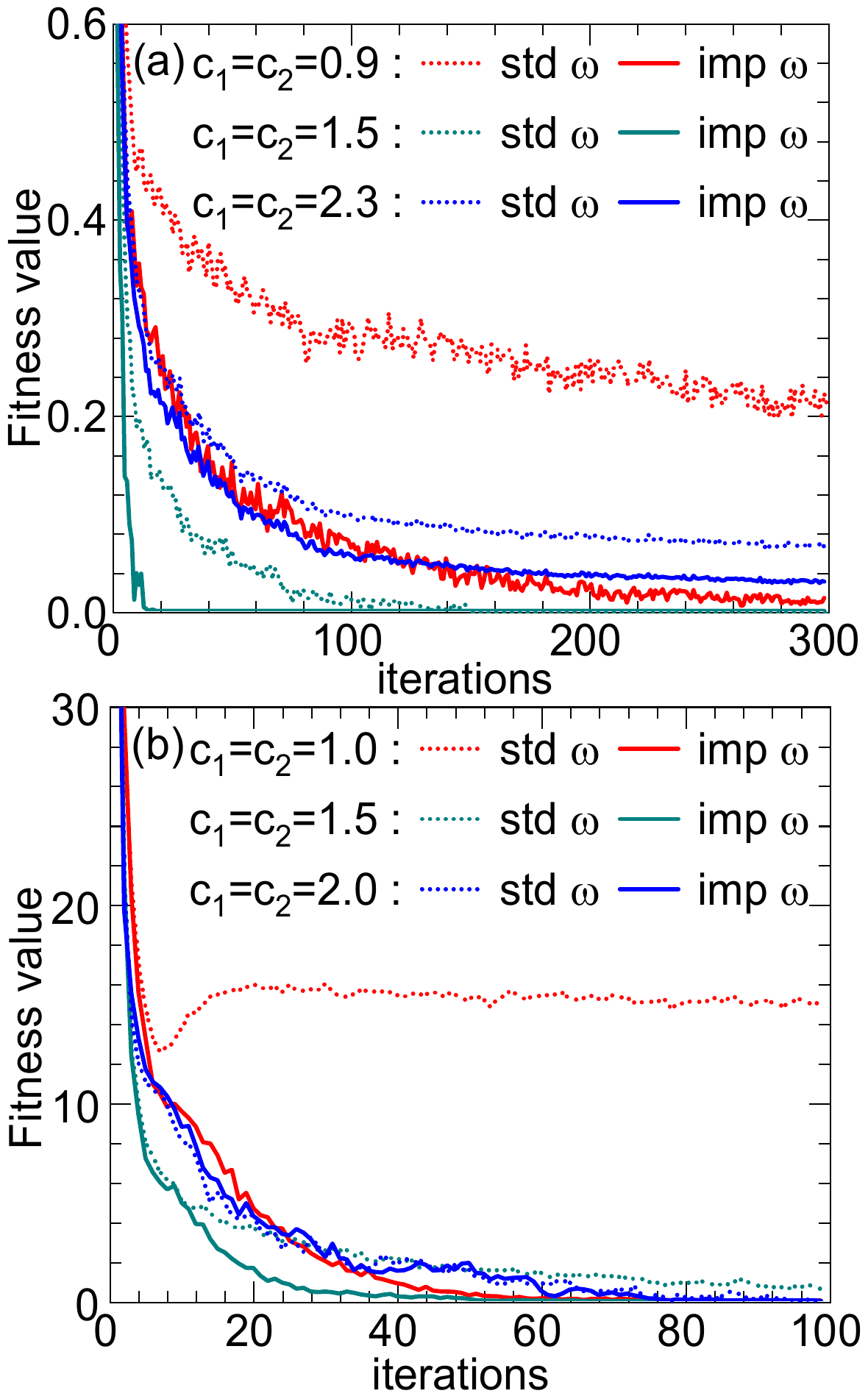}
\caption{The convergence process on the Griewank function (a) and Rastrigrin function (b) using PSO with improved $\omega$ or with standard $\omega$ where the learning factors ($c_{1}$ and $c_{2}$) of standard PSO are used.}
\label{inertia}
\end{figure}
We will now investigate the superiority of our improved PSO over standard PSO on two typical benchmark functions, namely the Griewank function and Rastrigrin function. As we have proposed the improved version for both the learning factors and inertia weight in our improved PSO, it is necessary to use the controlled variable method to separately investigate the performance improvement for standard PSO when including the learning factors or inertia weight individually.
\par
In Fig.\ref{learningfactors}, we compare the convergence process on the Griewank function [\ref{learningfactors}(a)] and Rastrigrin function [\ref{learningfactors}(b)] using PSO with improved $c_1$ and $c_{2}$ or with standard $c_{1}$ and $c_{2}$, where the inertia weight of standard PSO is used. Obviously, for these two typical benchmark functions, the convergence process of PSO using our improved version of $c_{1}$ and $c_{2}$ converges faster than that using $c_{1}$ and $c_{2}$ of standard PSO in a wide range of parameter space, indicating the superiority of our improved PSO over standard PSO regarding the learning factors.
\par
Furthermore, we compare the convergence process on the Griewank function and Rastrigrin function in Fig.\ref{inertia}(a) and \ref{inertia}(b), respectively, using PSO with the improved version of $\omega$ or with the standard version of $\omega$, where the learning factors of standard PSO are used. Similarly, our improved version of $\omega$ exhibits higher performance than the standard version of $\omega$ in wide range of $c_{1}$ and $c_{2}$ for these two typical benchmark functions.
\par
Thus, for both learning factors and inertia weight, our improved versions exhibit a higher performance than the standard PSO. It is necessary to mention that the combination of these two kinds of improved hyperparameters converges faster than the inclusion of learning factors or inertia weight individually.
\subsection{The application of improved PSO in the Hubbard model}
\begin{figure}[htbp]
\centering
\includegraphics[width=0.4\textwidth,height=0.68\textwidth]{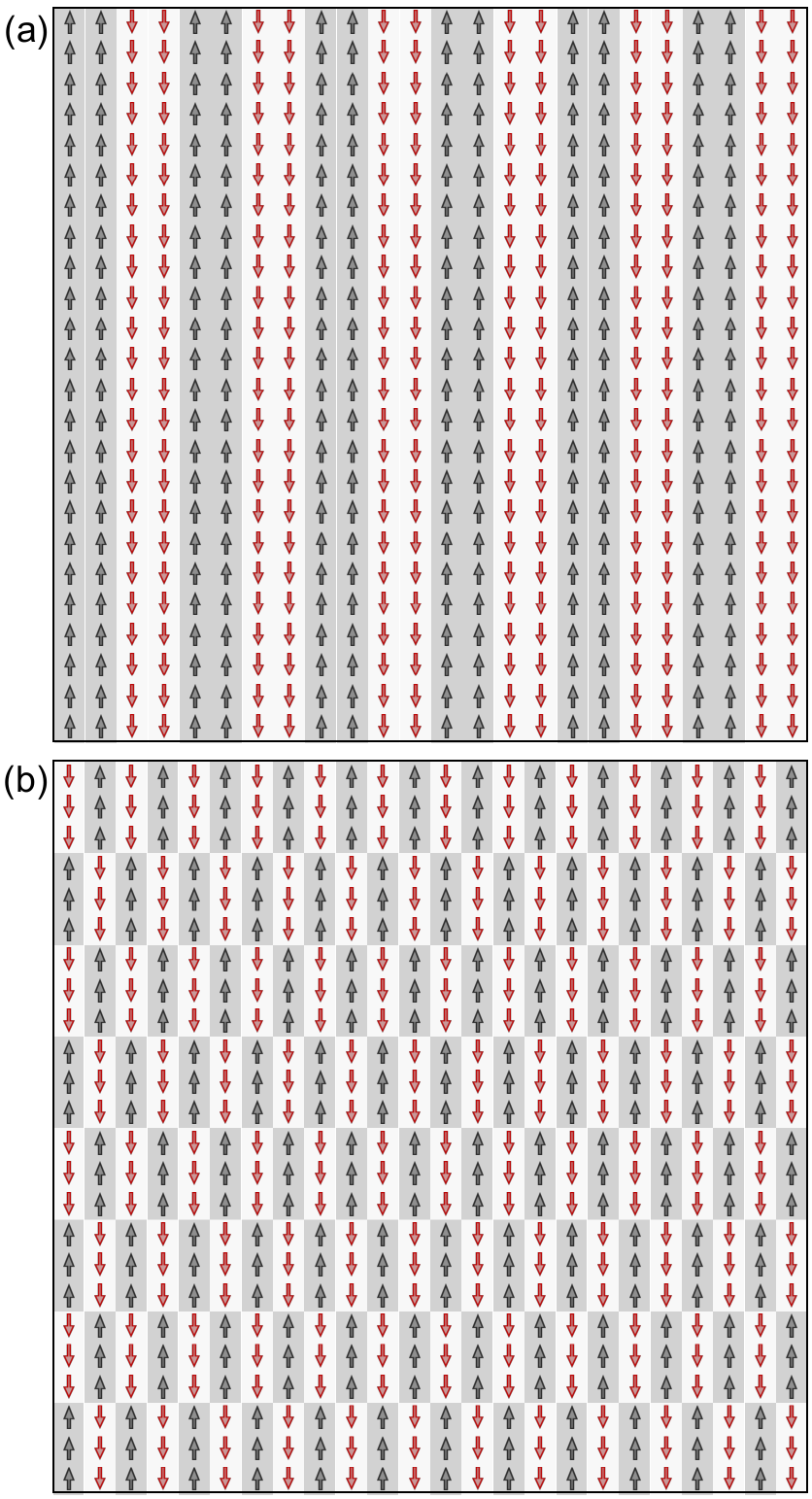}
\caption{The DASF and TAF patterns found by our improved PSO on the real-space square lattice with a finite size of 24$\times$24, where $t_{2}/t_{1}=0.99$ and $U/t_{1}=3.4$ in (a) while $t_{2}/t_{1}=0.765$ and $U/t_{1}=4.43$ in (b).}
\label{patterns}
\end{figure}
\begin{figure}[htbp]
\centering
\includegraphics[width=0.44\textwidth,height=0.74\textwidth]{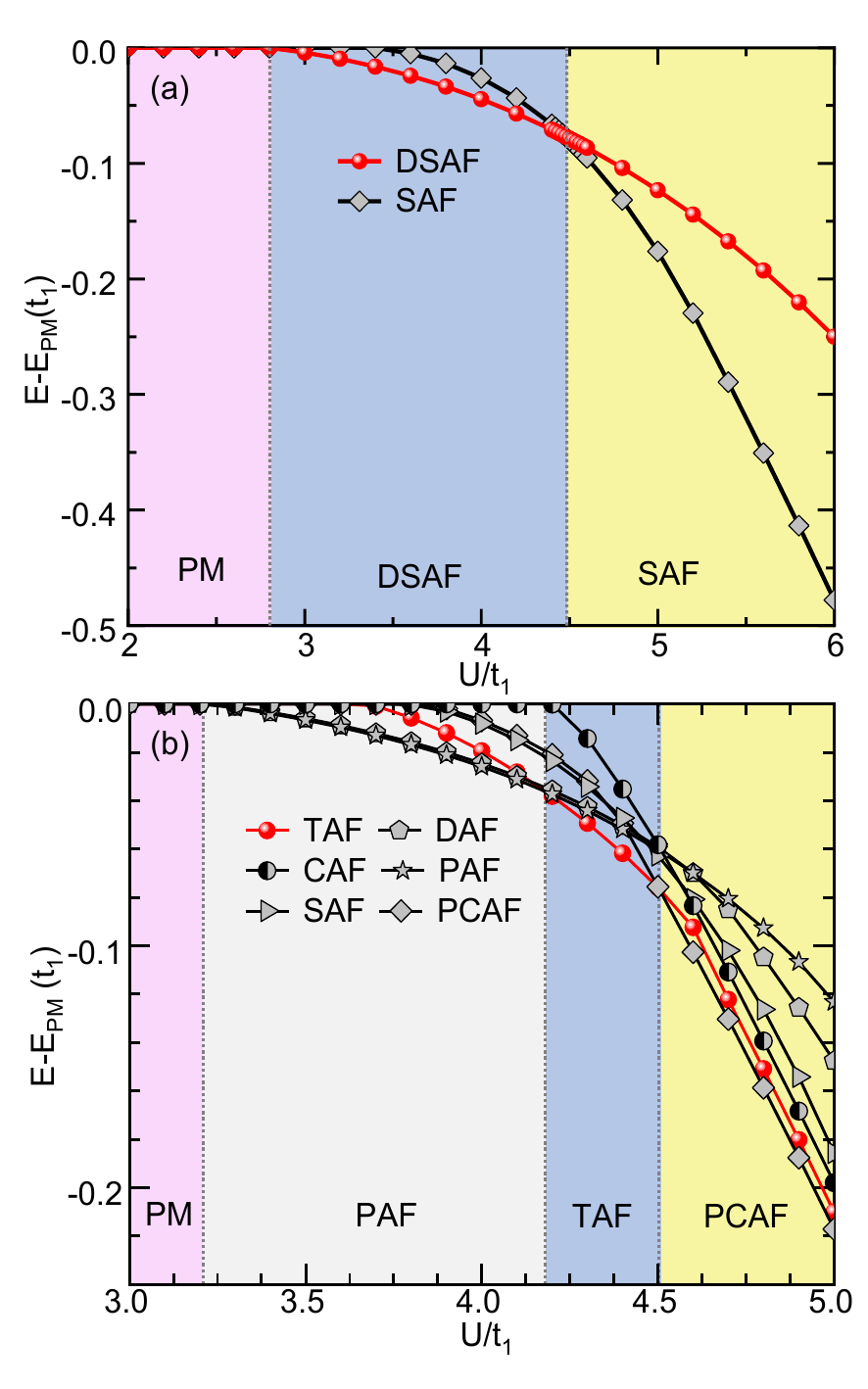}
\caption{(a) The free energies of SAF and DSAF as functions of the on-site Coulomb interaction $U$ at $t_{2}/t_{1}=0.99$, where the energy of the paramagnetic state is set to zero. (b) The free energies of CAF, DAF, TAF, PCAF, PAF and SAF as functions of U at $t_{2}/t_{1}=0.765$, where the energy of the paramagnetic state is set to zero.}
\label{energies-states}
\end{figure}
We then employ our improved PSO to search for new magnetic ground states of the Hubbard model on the real-space square lattice with a finite size of 24$\times$24 based on the mean-field approximation. Periodical boundary condition is used. Since particles, namely the magnetic configurations, will ultimately converge to the position where the free energy is minimized regardless of their initial positions within the framework of this algorithm, we initialize the magnetic configurations randomly and update them iteratively following the flowchart as shown in Fig.\ref{flowchart}. Notably, two new magnetic states, namely DSAF and TAF, are found apart from the existing states of CAF, DAF, PAF, PCAF, and SAF, where DSAF is in the region of weak geometrical frustration while TAF locates at strong geometrical frustration region. Interestingly, a Hubbard model with only the nearest and next-nearest neighbour hoppings can favor TAF, whereas $J_1$,$J_2$,$J_3$, and $K$ are needed for a Heisenberg model~\cite{glasbrenner2015effect}. The specific configurations of DSAF and TAF are shown in Fig.\ref{patterns}(a) and \ref{patterns}(b), respectively. It is necessary to mention that the calculations here are performed on a finite size with only 24$\times$24, which may be affected by finite-size effects.

To confirm that these two new magnetic states are not a result of a finite-size effect, we further perform mean-field calculations in the thermodynamical limit. Figure \ref{energies-states} shows the comparison of free energies between DSAF and its corresponding competing states \ref{energies-states}(a), as well as between TAF and its corresponding competing states \ref{energies-states}(b). Obviously, DSAF occurs in the region between the paramagnetic state (PM) and SAF, while TAF exists in the region between PAF and PCAF. Thus, the presence of our two new found magnetic states is confirmed in the thermodynamical limit, which is not a result of a finite-size effect.

\begin{figure}
\centering
\includegraphics[width=0.48\textwidth,height=0.25\textwidth]{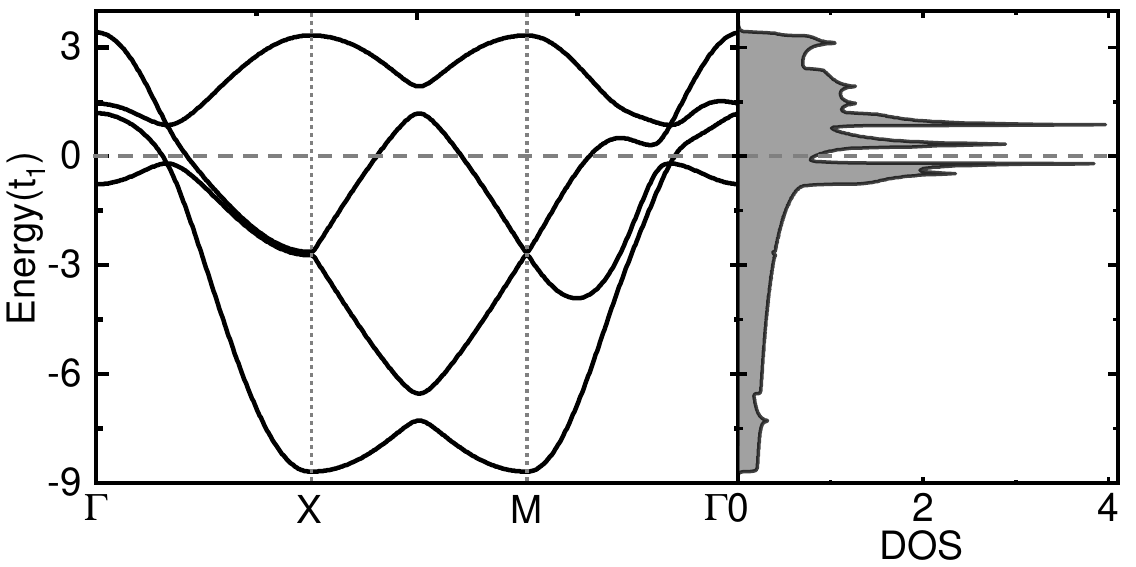}
\caption{Band structure and corresponding DOS for the DSAF state at $t_{2}/t_{1}=0.99$ and $U/t_{1}=3.6$.}
\label{DSAF-DOS-BS}
\end{figure}

\begin{figure}
\centering
\includegraphics[width=0.48\textwidth,height=0.25\textwidth]{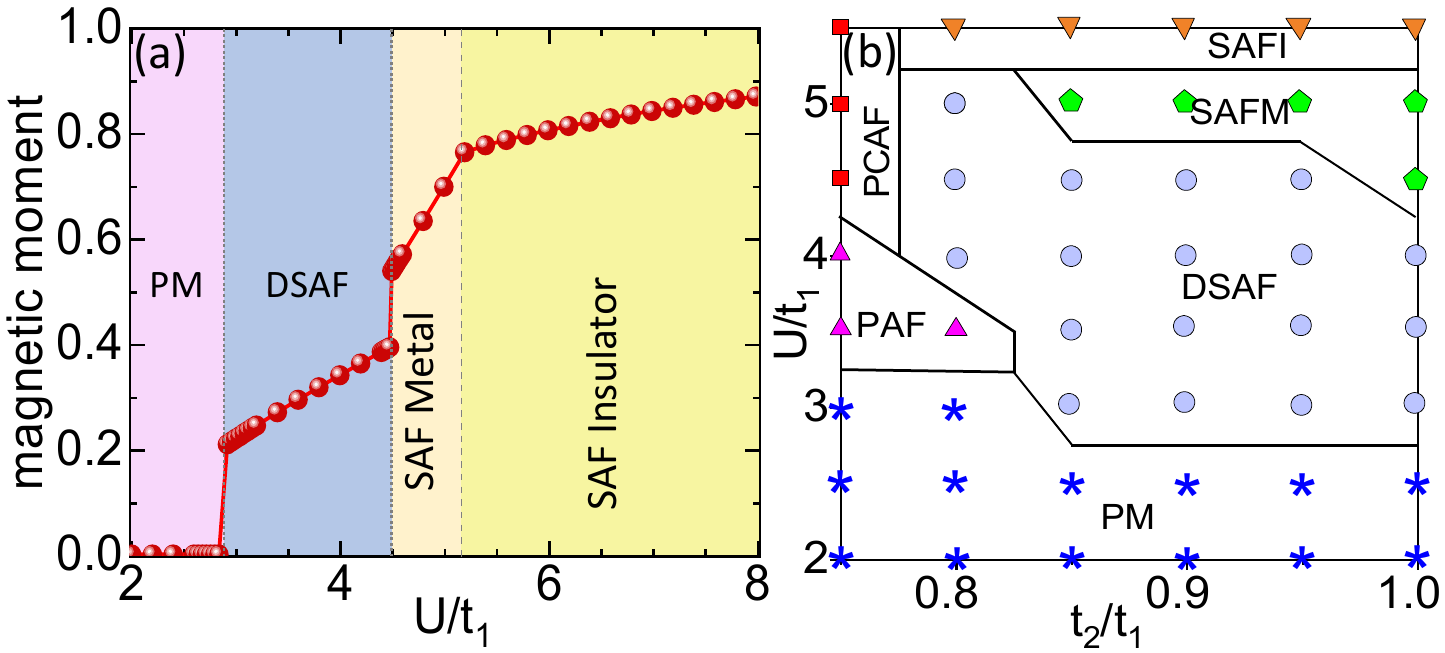}
\caption{(a) The magnetic moment as a function of the on-site Coulomb interaction $U$ at $t_{2}/t_{1}=0.99$. (b) A schematic phase diagram of DSAF in the region of $0.75<t_{2}/t_{1}<1$ and $2<U/t_{1}<5.5$, in which the star, square, triangle, inverted triangle, pentagon, and circle denote PM, PCAF, PAF, SAFI, SAFM, and DSAF, respectively.}
\label{magnet}
\end{figure}

Considering that DSAF occurs at $2.8<U/t_{1}<4.5$ ($0.23<U/W<0.36$, where $W$ is the bandwidth) for $t_{2}/t_{1}=0.99$, where geometrical frustration is released and the electronic correlation is relatively weak, this state may be reliable within mean-field approximation. Thus, we investigate the band structure and density of states (DOS) for DSAF in Fig.\ref{DSAF-DOS-BS}. As can be seen, DSAF is metallic. Furthermore, we demonstrate the magnetic moments of DSAF and its corresponding competing states as a function of the on-site Coulomb interaction $U$ at $t_{2}/t_{1}=0.99$ in Fig.\ref{magnet} (a). The sudden enhancements of the magnetic moments from PM to DSAF and from DSAF to SAF suggest first-order phase transitions. The kink occurring in the magnetic moment for the case of SAF is a result of phase transition from metallic SAF to insulating SAF, consistent with previous study~\cite{yu2010collinear}. We also present a schematic phase diagram of DSAF in Fig.\ref{magnet} (b). Obviously, DSAF becomes the ground state within a broad region for the magnetic states we considered.

\section{discussion}
Here, we have proposed a new idea about how to search for magnetic ground states of the Hubbard model using PSO based on mean-field approximation by viewing magnetic configurations as particles. We have found two new magnetic ground states in the Hubbard model with the nearest and next-nearest neighbour hoppings by employing our improved version of PSO. Although the PSO calculations in this paper are based on mean-field approximation, PSO is a very powerful algorithm that can be effectively combined with higher-level approximations to handle the Hubbard model, such as Hubbard-\uppercase\expandafter{\romannumeral1} approximation~\cite{hubbard1-1963electron}, Hubbard-\uppercase\expandafter{\romannumeral3} approximation~\cite{hubbard3-1964electron}, projective operator approximation~\cite{roth1969electron,fan2018projective}, coherent potential approximation~\cite{soven1967coherent,velicky1968single,xu2016gate}, dynamical mean-field theory~\cite{georges1992hubbard,georges1996dynamical}, etc.

Twisting often leads to the emergence of exotic magnetic phases in layered materials, such as the coexistence of interlayer ferromagnetic and interlayer antiferromagnetic states in twisted bilayer CrI$_3$~\cite{xu2022coexisting}. However, enumerating all the possible magnetic states in twisted layered materials becomes challenging due to the large number of sublattices in the supercell. Fortunately, PSO offers an excellent platform to automatically search for the magnetic ground state without the need for manual preparation of magnetic configurations. Therefore, the application of PSO in studying the magnetism of twisted systems is expected to be very interesting.

By using PSO, we have discovered two new magnetic states in the Hubbard model we investigated, including DSAF and TAF. Although a double-Q coplanar spin-vortex crystal phase is proposed when $t_2\approx t_1$ and $U/W > 0.31$ ($U/t_1 > 3.8$)~\cite{huang2020antiferromagnetic}, our DSAF occurs at the region with a weaker correlation interaction of $U/W > 0.23$ ($U/t_1 > 2.8$). Besides, since the geometrical frustration is released and the electronic correlation is weak, mean field approximation is reliable, as evidenced by qualitative consistencies between solutions from mean field approximation~\cite{ruan2021uncovering} and VCA\cite{nevidomskyy2008magnetism} where SAF starts to appear at $U/t_1$ less than 4 if DSAF is not taken into account, indicating that quantum fluctuations completely ignored in mean field approximation are suppressed in the large $t_2/t_1$ and weak $U/W$ region. Since the free energy of DSAF is significantly lower than that of other states in the DSAF phase, it is expectable that the stable DSAF solution should survive against the quantum fluctuations as even the metastable SAF solution has already survived against the quantum fluctuations\cite{nevidomskyy2008magnetism}. Additionally, we find that a prominent peak is present in the vicinity of $(\pi/2, 0)$ for the Pauli susceptibility in the $t_2 \approx t_1$ region, indicating a tendency towards DSAF if perturbation, like the on-site Coulomb interaction, is switched on, further supporting our finding of DSAF in the certain region. Thus, DSAF is most probably a genuine ground state in a sufficient large parameter space for this Hubbard model even if both non-collinear magnetism and quantum fluctuations are considered. In contrast, TAF may be controversial since it occurs in the region of strong geometric frustration where macroscopic degeneracies dramatically affect the magnetism of the one-band Hubbard model. As a result, various competing states are proposed in this region, such as PCAF~\cite{mizusaki2006gapless,yamada2013magnetic}, DAF~\cite{ruan2021uncovering}, PAF~\cite{ruan2021uncovering}, as well as nonmagnetic insulating state~\cite{mizusaki2006gapless}. Thus, it is interesting to use a more sophisticate method than mean-field approximation to confirm the stability of TAF over its corresponding competing states.

Compared to standard PSO, our improved PSO exhibits a higher performance because the search direction of particles in the standard PSO is so random that it lacks sufficient guidance. In contrast, by proposing an improved version of the learning factors and inertia weight, particles have their own characters to recognize the searching direction in our improved PSO, which significantly save the time of particles wandering around local optimum. Thus, our improved PSO converges faster than standard PSO. Our work provides valuable insights into how to improve PSO by modifying the learning factors and inertia weight and demonstrates its effectiveness in searching new states of matters.

\section{conclusion}
In conclusion, we have proposed an improved version of PSO, which converges to global optimum faster than standard PSO. By employing this improved PSO to search for new magnetic states in the Hubbard model with the nearest and next-nearest neighbour hoppings on the real-space square lattice with a finite size of 24$\times$24 based on the mean-field approximation, two new magnetic states, including DSAF and TAF, are found. The presence of these two states in this Hubbard model is further confirmed by mean-field calculations in the thermodynamical limit, where the band structure, DOS, and magnetic moment of DSAF are also present.
\section{Acknowledgement}
This work is supported by National Natural Science Foundation of China (Grants No. 12274324, No. 12004283) and Shanghai Science and Technology Commission (Grant No. 21JC405700).

\bibliography{PSO_references}
\end{document}